\documentclass{ws-ijmpb}
\usepackage{graphicx}
\usepackage{amsmath,amssymb,amsfonts}
\usepackage{bm}

\def\be{\begin{equation}} \def\ee{\end{equation}}
\def\bea{\begin{eqnarray}} \def\eea{\end{eqnarray}}
\def\nn{\nonumber}

\newcommand{\Ham}{\mathcal{H}}
\newcommand{\hbT}{\hat{\bm{T}}}
\newcommand{\hT}{\hat{T}}
\newcommand{\hbS}{\hat{\bm{S}}}

\begin{document}

\markboth{W.-C. Lee, W. Lv, and H. Z. Arham}
{Elementary Excitations due to Orbital Degrees of Freedom in Iron Based Superconductors}

%
\catchline{}{}{}{}{}
%

\title{Elementary Excitations due to Orbital Degrees of Freedom in Iron Based Superconductors}

\author{WEI-CHENG LEE}

\address{Department of Physics, University of Illinois, 1110 W. Green Street\\ Urbana, Illinois 61801, United States of America
\\
leewc@illinois.edu}

\author{WEICHENG LV}

\address{Department of Physics and Astronomy, University of Tennessee\\ Knoxville, Tennessee 37996, United States of America \\ Materials Science and Technology Division, Oak Ridge National Laboratory\\ Oak Ridge, Tennessee 37831, United States of America}

\author{HAMOOD Z. ARHAM}

\address{Department of Physics, University of Illinois, 1110 W. Green Street\\ Urbana, Illinois 61801, United States of America}

\maketitle

\begin{abstract}
One central issue under intense debate in the study of the iron based superconductors is the origin of the structural phase transition that changes the crystal lattice symmetry from tetragonal to orthorhombic. This structural phase transition, occurring universally in almost every family of the iron-based superconductors,
breaks the lattice $C_4$ rotational symmetry and results in an
anisotropy in a number of physical properties. Due to the unique topology of the Fermi surface, both orbital- and spin-based scenarios have been proposed as the driving force.
In this review, we focus on theories from the orbital-based scenario and discuss several related experiments.
It is pointed out that although both scenarios lead to the same macroscopic phases and are not distinguishable in bulk measurements of the thermodynamic properties,
the elementary excitations could be fundamentally different, and provide us with the possibility to resolve this long-standing debate between orbital- and spin-based theories.
\end{abstract}


\section{Introduction}
\label{sec:intro}
Understanding the nature of the excitations in both the normal and the superconducting states has been a heavily studied research topic 
in high-temperature superconductors, for it might hold the key to the pairing mechanism for unconventional superconductivity, the ``Holy Grail'' of the field. It is no doubt that the spin physics plays a crucial role in this aspect due to the presence of
some magnetically ordered states in the parent compounds.
After the doping is introduced, the superconductivity emerges with the disappearance of the `parent' magnetic order.
Since this sort of phase diagram is so universal, it is reasonable to believe that the spin fluctuations originating from the `parent' magnetic order
persist even as the long-ranged order is killed, and they might take a part in the formation of the Cooper pairs.
A number of experiments support the ubiquity of the spin fluctuations among cuprates, heavy fermions, and iron pnictides.\cite{scalapino2012}

The new high temperature superconductor discovered in 2008, the iron pnictide, has received a tremendous amount of attention in the past few years.\cite{Paglione2010,Johnston2010}
At first sight, this category of materials seems to be incompatible with superconductivity.
The key element, Fe$^{2+}$, has six electrons in its 3$d$ orbitals, and Hund's coupling should favor ferromagnetism (parallel spins) instead of spin-singlet superconductivity (anti-parallel spins).
Indeed there were some theoretical proposals of the iron pnictide being a spin-triplet $p$-wave superconductor,\cite{Lee2008} but this was ruled out very quickly by
experiments.
The hybridizations between the $d$ orbitals of the Fe atom and the $p$ orbitals of the As atom result in a strong suppression of the effect of Hund's coupling.\cite{Wu2008}
Consequently, the magnetic moment of the Fe atom is surprisingly small, which prevents the occurrences of ferromagnetism as well as spin-triplet superconductivity.
Experimentally it has been found that the Fe atom has a magnetic moment ranging from 0.5 to 3.3 $\mu_B$ in different families of the iron pnictides, depending on the details of their
electronic structures. This sensitivity of the Fe moment on the electronic structures reflects the hybridization effect discussed above.

Another crucial consequence due to the strong hybridization is the multiorbital nature.
As seen in density functional theory (DFT) calculations,\cite{singh2008,haule2008,cao2008}
the Fermi surface of the iron pnictides dissects into different pockets centered around high symmetry points of the Brillouin zone $\Gamma$ and $M$.
A closer look into the band structures makes it obvious that the eigenstates on different parts of the Fermi surface have very different
weights in the Fe $d$ orbitals.
In fact, it is just this multiorbital nature that makes the iron pnictides distinct from the cuprates.

\begin{figure}[t]
  \centering
  \includegraphics[width=\textwidth]{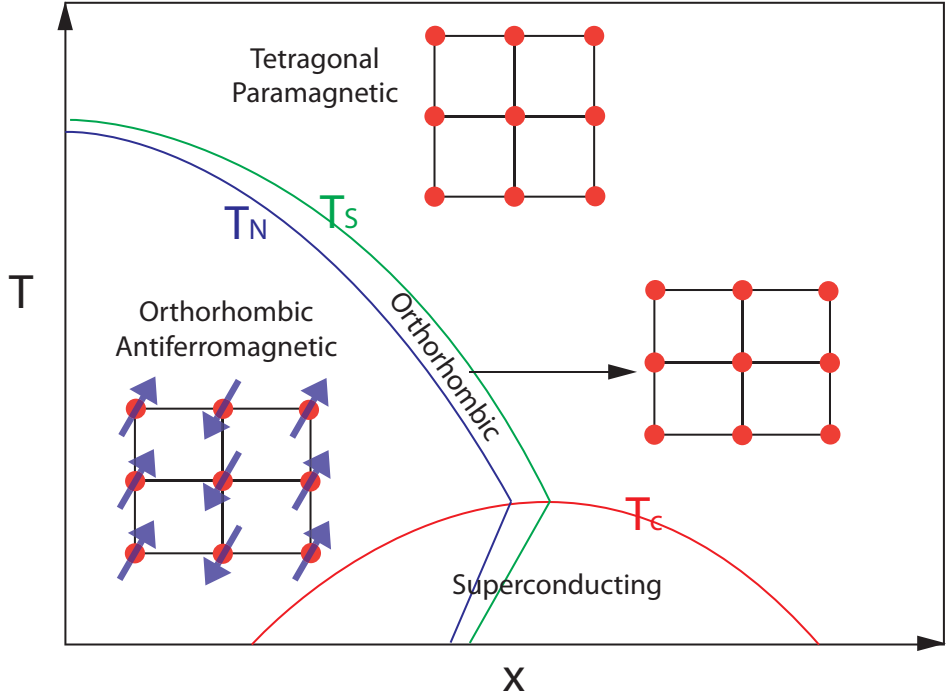}
  \caption{The generic phase diagram of the iron-based superconductors as a function of doping $x$ and temperature $T$. We use $T_S$, $T_N$, and $T_C$ to label the structural, magnetic, and superconducting phase transition temperatures, respectively. The corresponding crystal and magnetic structures of each phase are displayed.}
  \label{fig:phase}
\end{figure}

Fig.~\ref{fig:phase} shows a generic phase diagram of the iron-based superconductors. Let us first discuss the part that is remarkably similar to that of the cuprates.
The parent compound has a stripe-like antiferromagnetic (AFM) order with the ordering wave vector $\bm{Q} = (\pi,0)$, while superconductivity appears as the long-ranged AFM order
is suppressed by either doping or pressure.
However, a new ingredient that is not universally present in the cuprates is the emergence of a structural phase transition,
across which the crystal symmetry goes from tetragonal to orthorhombic.
The structural phase transition spontaneously breaks the lattice $C_4$ rotational symmetry and always occurs at a temperature ($T_S$)
which is the same as or higher than the magnetic transition temperature ($T_N$).
This unusual inequality ($T_S\geq T_N$) has motivated theorists to propose either orbital or spin fluctuations as the effective cause.
The difficulty to resolve this controversy is that both theories end up with the same macroscopic states breaking exactly the same $C_4$ symmetries,
so that they are indistinguishable in most bulk measurements of the thermodynamic properties.

In this review, we focus on the orbital physics emerging from this multiorbital nature which is new and unique in the iron pnictides.
It is shown that the elementary excitations of the orbital fluctuations is fundamentally different from those of the spin fluctuations, which might give us a chance to
resolve the nature of the quantum fluctuations from various types of spectroscopies.
This paper is organized as follows. Sec.~\ref{sec:orbital} introduces the ferro-orbital order of Fe $d_{xz}$ and $d_{yz}$ orbitals as the effective cause of the structural phase transition. In particular, we will show how this orbital order arises from both the strong-coupling and the weak-coupling limit.
In Sec.~\ref{sec:excitation}, the elementary excitations from the quantum fluctuations associated with the ferro-orbital order are discussed, and the
experiments that might have detected these orbital fluctuations are summarized in Sec. 4.
Finally, a concluding remark is given in Sec. 5.

\section{Orbital Order}
\label{sec:orbital}
\subsection{Structural phase transition and nematic order}
The ground state manifold of the magnetically ordered phase of the iron-based superconductors is $O(3) \times Z_2$, where $O(3)$ refers to the arbitrary direction of the ordered moment in spin space, whereas $Z_2$ represents the two-fold degeneracy of the ordering wave vector $(\pi,0)$ and $(0,\pi)$. This magnetic order not only breaks the usual $O(3)$ spin rotational symmetry, but also breaks the $C_4$ rotational symmetry of the tetragonal lattice. From a symmetry argument, accompanying the onset of the magnetic order, a structural phase transition could occur and
reduce the crystal symmetry from tetragonal to orthorhombic. Furthermore, a more intriguing possibility is that the $C_4$ symmetry breaking can arise even without breaking the spin $O(3)$ symmetry. Indeed, experimentally the structural phase transition temperature $T_S$ is higher than the N\'eel temperature $T_N$ in many iron-based superconductors. The phase sandwiched between $T_S$ and $T_N$ is usually called the nematic phase, which proliferates in many strongly correlated systems.\cite{fradkin2010} In its original meaning, the nematic order is characterized by a state that breaks a \emph{continuous} rotational symmetry but remains invariant under other symmetry operations. However, in real materials, the underlying symmetry of the system becomes discrete due to the presence of the lattice. So the nematic order only breaks a \emph{discrete} rotational symmetry, in particular the $C_4$ rotational symmetry of the two-dimensional square lattice considered here.

Understanding the origin of the $C_4$ symmetry breaking associated with the structural phase transition is a fundamental question in the iron-based superconductors. Recently a lot of experimental efforts have been made to uncover the electronic anisotropy in this nematic phase. Evidences from transport measurement,\cite{Chu2010,Tanatar,Dusza2012} angle-resolved photoemission spectroscopy (ARPES),\cite{Yi2011} and inelastic neutron scattering\cite{Harriger2011,Harriger2012} all suggest that strong electronic anisotropy exist above $T_N$ (for a review, see Ref.~\refcite{Fisher}), and in some cases, it even persists above $T_S$ with or without uniaxial strains\cite{Matsuda}. We note here that the orthorhombic distortion is very small, less than 1\% of the original lattice. So it is difficult to ascribe the strong electronic anisotropy to lattice distortions. Therefore it is generally agreed that the symmetry breaking has an electronic origin, which is usually termed as the electron nematic order.\cite{oganesyan2001}

In this section, we focus on theories in which the orbital order is an effective cause of the electron nematicity.\cite{Lv2009,Lee2009,Turner2009} As mentioned in Sec.~\ref{sec:intro}, iron-based superconductors are intrinsically multiorbital systems, and DFT calculations\cite{cao2008,graser2009} show that all five Fe 3$d$ orbitals have significant weights on the Fermi surface.
Among five $d$ orbitals, two of them, $d_{xz}$ and $d_{yz}$, are special from the symmetry aspect. 
Each of them has only the $C_2$ symmetry, but the system retains the $C_4$ symmetry if they are degenerate.
Moreover, they make dominant contributions on the Fermi surface compared to the other three orbitals.
Therefore the minimal model for iron based superconductors is a two-orbital model with the $d_{xz}$ and $d_{yz}$ orbitals.\cite{raghu2008}
In such a model, we can define the ferro-orbital order parameter as the difference between the occupation numbers of the two orbitals, $m =n_{xz} - n_{yz}$. A non-zero $m$ lifts the orbital degeneracy and changes sign under the $C_4$ rotation. Therefore it can be identified as the electron nematic order parameter, which gives rise to the structural phase transition.

What is the microscopic mechanism driving the orbital order? Before answering this question, we need to address the important issue regarding the strength of electron correlations. Because various experiments have characterized the iron-based superconductors as intermediately correlated metals,\cite{Qazilbash2009,Yang2009} both strong-coupling\cite{Dai2012} and weak-coupling\cite{Hirschfeld2011} theories have their own merits. So we will use arguments from both aspects to show how the orbital order arises in Sec.~\ref{sec:strong} and \ref{sec:weak}, respectively.

Here we briefly mention the spin-based scenario for the structural phase transition. In this line of thinking, the nematic order is induced by magnetic fluctuations,
termed as the spin nematic order. These theories can be formulated either from a strong-coupling $J_1$-$J_2$ Heisenberg model,\cite{Fang2008,Xu2008} or from a weak-coupling model consisting of electron and hole pockets.\cite{Fernandes2012} Although starting from the opposite limits, both approaches end up with the same effective Landau-Ginzberg model, in which the nematic order can be identified as a $Z_2$ order related to magnetic fluctuations centered at $(\pi,0)$ and $(0,\pi)$.
Once the nematic order sets in, the magnetic fluctuations will be enhanced at a preferable wavevector without the onset of the long-ranged magnetic order. This approach naturally explains the proximity of structural and magnetic transitions. From a symmetry point of view, both the orbital order and spin nematic order break the same $C_4$ rotational symmetry. Thus they are allowed to couple to each other directly. Namely, the onset of one order parameter will induce the other. It is therefore very difficult to distinguish their roles in causing the electron nematic order. However, a key point to be emphasized is that orbital order gives a new energy scale that is distinct from spin physics.
As we will show in Sec.~\ref{sec:excitation} and \ref{sec:expt}, orbital order leads to a new type of elementary excitations that has strong support from experiments.

\subsection{Orbital order from strong coupling}
\label{sec:strong}
The idea of orbital order in the insulating limit is pioneered by the seminal work of Kugel and Khomskii.\cite{Kugel1982} We will apply their approach to the iron-based superconductors, in particular a two-orbital model consisting of $d_{xz}$ and $d_{yz}$ orbitals at quarter filling. On each site there is one electron occupying any state that is a linear combination of $d_{xz}$ and $d_{yz}$. Formally, such a state is written as
\be
    \vert \theta \rangle = \cos\left(\frac{\theta}{2} \right) \vert xz \rangle +
    \sin\left(\frac{\theta}{2} \right) \vert yz \rangle.
\ee
Due to the two-fold degeneracy, we can introduce an orbital pseudospin operator $\hbT$ with $T=1/2$ to represent the two orbital states, where
\be
    \vert xz \rangle = \left(\begin{array}{c}1\\ 0 \end{array} \right), \quad
    \vert yz \rangle = \left(\begin{array}{c}0\\ 1 \end{array} \right).
\ee

According to Kugel and Khomskii,\cite{Kugel1982} this orbital degeneracy can be lifted by three types of interactions in a lattice system. The first one is the quadrupole-quadrupole interaction arising from the quadrupole moment of the 3$d$ orbitals. In terms of orbital pseudospins $\hbT$, it takes the form of
\be
    \Ham_{qu} = \sum_{ij} J^z_{ij} \hT_i^z \hT_j^z + J^x_{ij} \hT_i^x \hT_j^x + J^{zx}_{ij} \left( \hT_i^z \hT_j^x + \hT_i^x \hT_j^z \right).
\label{eq:heisenberg}
\ee
Depending on the values of the exchange constants $J$, different orbitally ordered states can be stabilized. If $J^z_{ij}$ is negative and the leading energy scale of the system, ferro-orbital order will develop at low temperature, where $m = \langle \hT_i^z \rangle \neq 0$.

The second one is the electron-phonon coupling,
\bea
    \Ham_{ep} & = & \sum_{k} \omega_k a_k^\dagger a_k - \frac{1}{\sqrt{N}} \sum_{k,i} e^{-ikr_i} g_k \left(a_k+a_k^\dagger \right) \hT_i^z
     + \frac{C}{2} \delta^2 - \frac{g_0}{N} \sum_i \delta \hT_i^z.
\eea
The fist term represents free phonons with frequency $\omega_k$, while the second term describes the electron-phonon interaction, with $g_k$ being the coupling strength. We have separated the contributions from homogeneous distortions in the last two terms. Here $\delta$ is the orthorhombic distortion; $C$ is the elastic constant; and $g_0$ is the coupling strength. We can formally integrate out the lattice variables and derive an effective model of pseudospins,
\be
    \Ham\left(\hbT_i, \hbT_j \right) = \sum_{ij} \tilde{J}_{ij} \hT^z_i \hT^z_j
\label{eq:ising}
\ee
where the exchange constants
\be
    \tilde{J}_{ij} = - \frac{g_0^2}{C} - \sum_{k} \frac{g_k^2}{\omega_k} e^{ik\left(r_i - r_j \right)}.
\ee
We note that only the orthorhombic distortion that couples to $\hT^z$ is considered here. As a result, the effective Hamiltonian of Eq.~(\ref{eq:ising}) takes the form of an Ising model. In reality, there is another phonon mode that couples to $\hT^x$. Inclusion of this coupling will produce terms of the form $\hT_i^x \hT_j^x$ and $\hT_i^x \hT_j^z$, resulting in an effective Hamiltonian similar to the form of Eq.~(\ref{eq:heisenberg}). Nevertheless, since ferro-orbital order and the accompanying orthorhombic distortion are the leading instability in our system, it is sufficient to focus only on $\hT^z$. Similar to the quadrupole-quadrupole interaction, electron-phonon interaction also leads to a ferro-orbitally ordered phase at low temperature.

The last type of interaction is the exchange interaction. Unlike the first two interactions, which can be expressed exclusively by orbital pseudospins $\hbT$, the exchange interaction also involves the spin degrees of freedom. In many systems, it is the dominant energy scale for orbital and spin orders. Formally this spin-orbital superexchange model can be derived from a second-order strong-coupling expansion of a multiorbital Hubbard model. Considering the virtual hopping process $d^1d^1 \rightarrow d^2d^0 \rightarrow d^1d^1$, we write down its general form along a given bond $(i,j)$,\cite{Kruger2009}
\be
    \Ham_{SO}^{(i,j)}  =   \sum_{\tau_i,\tau_j}\sum_{s_i,s_j} J_{\tau_i,\tau_j,s_i,s_j}^{(i,j)}
    A^{(i,j)}_{\tau_i,\tau_j}(\hbT_i, \hbT_j)
    \times B_{s_i,s_j}(\hbS_i, \hbS_j),
\ee
where $\hbS$ and $\hbT$ are the $S=1/2$ spin and the $T=1/2$ orbital pseudospin operators, respectively. Here we use ($\tau_i$, $\tau_j$) and ($s_i$, $s_j$) to label the orbital and spin quantum numbers of the intermediate high-energy state.

\begin{figure}[t]
  \centering
  \includegraphics[width=\textwidth]{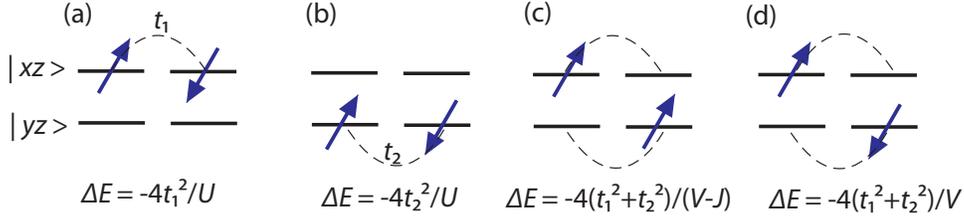}
  \caption{Illustration of the spin-orbital coupling. We consider virtual hopping processes along the $x$ direction. The intra-orbital hopping amplitudes of $d_{xz}$ and $d_{yz}$ are $t_1$ and $t_2$, respectively. The energy of each orbital and spin configuration can be calculated using second-order perturbation theory. $U$, $V$, and $J$ are intra- and inter-orbital Coulomb repulsion, and the Hund's coupling, respectively.}
  \label{fig:virtual}
\end{figure}

Instead of presenting a formal derivation of the spin-orbital model, here we use simple arguments from Goodenough\cite{Goodenough1955} and Kanamori\cite{Kanamori1959} to illustrate how spin and orbital orders are coupled to each other. As an example, let us consider the virtual hopping processes between nearest neighbors (NN) along the $x$ direction for a two-orbital model. We define the intra-orbital hopping amplitudes of the $d_{xz}$ and $d_{yz}$ orbitals as $t_1$ and $t_2$, respectively. The inter-orbital hopping vanishes from symmetry arguments. We need to consider three distinct orbital configurations as illustrated in Fig.~\ref{fig:virtual}. If both sites are occupied by the same orbitals [Fig.~\ref{fig:virtual}(a) and (b)], the exchange interaction is strongly antiferromagnetic (AFM). We can write down an effective spin-only Heisenberg model
$\Ham(\hbS_i,\hbS_j) = J \hbS_i \cdot \hbS_j$, where $J=J_{1a}=4t_1^2/U$ if $d_{xz}$ occupies both sites, while $J=J_{1b}=4t_2^2/U$ when we have $d_{yz}$ orbitals on both sites.
Here $U$ is the intra-orbital Coulomb repulsion.
In contrast, if different orbitals occupy the two sites [Fig.~\ref{fig:virtual}(c) and (d)], the exchange becomes weakly ferromagnetic (FM), with
\be
    \Ham\left( \bm{S}_i, \bm{S}_j\right) = \left(\frac{4\left(t_1^2+t_2^2\right)}{V} - \frac{4\left(t_1^2+t_2^2\right)}{V-J} \right)  \bm{S}_i \cdot \bm{S}_j \approx  - \frac{4\left(t_1^2+t_2^2\right)J}{V^2}  \bm{S}_i \cdot \bm{S}_j,
\label{eq:fm}
\ee
in the limit that the Hund coupling $J$ is much smaller than the inter-orbital Coulomb repulsion $V$. Consequently, different orbital configurations correspond to different spin configurations, and vice versa. Namely, the orbital and spin orders are strongly coupled. If we further assume that $t_1 \gg t_2$ according to the spatial anisotropy of the orbitals, it is found that the state with the $d_{xz}$ ferro-orbital order and AFM spin order will have the lowest energy. The same argument can be applied to NNs along the $y$ direction. Along the diagonal direction between next nearest neighbors (NNN), the inter-orbital hopping no longer vanishes and there is no dominant hopping amplitude. Therefore we may make the simplification that the exchange constant is orbitally independent and equals $J_2$ no matter which orbitals are occupied.

\begin{figure}[t]
  \centering
  \includegraphics[width=\textwidth]{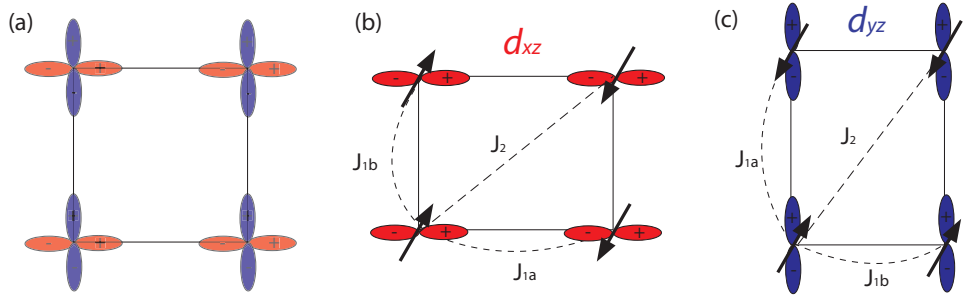}
  \caption{Orbital order and the resulting magnetic order with the anisotropic exchange constants. (a) At high temperature, $d_{xz}$ and $d_{yz}$ orbitals are equally populated. (b) The $d_{xz}$ orbital order and the $\bm{Q} = (\pi,0)$ AFM order. (c) The $d_{yz}$ orbital order and the $\bm{Q} = (0,\pi)$ AFM order. $J_{1a}$, $J_{1b}$, and $J_2$ are the exchange constants between NNs and NNNs.}
  \label{fig:oo_strong}
\end{figure}

As a result, we can write down the following spin-orbital model to describe the structural-magnetic phase transitions in iron-based superconductors,\cite{Lv2009,Chen2009}
\bea
\Ham_{SO} & = & \sum_{i} \left[ J_{1a} \left(\hT_i^z + \frac{1}{2}\right) \left(\hT_{i+\hat{x}}^z + \frac{1}{2}\right) + J_{1b} \left(\hT_i^z - \frac{1}{2}\right) \left(\hT_{i+\hat{x}}^z - \frac{1}{2}\right) \right] \hbS_i \cdot \hbS_{i+\hat{x}} \nn \\
& & \, + \sum_{i} \left[ J_{1b} \left(\hT_i^z + \frac{1}{2}\right) \left(\hT_{i+\hat{y}}^z + \frac{1}{2}\right) + J_{1a} \left(\hT_i^z - \frac{1}{2}\right) \left(\hT_{i+\hat{y}}^z - \frac{1}{2}\right) \right] \hbS_i \cdot \hbS_{i+\hat{y}} \nn \\
& & \, + \sum_{\langle \langle i,j \rangle \rangle} J_2 \hbS_i \cdot \hbS_j.
\eea
We have ignored the weak FM exchange from Eq.~(\ref{eq:fm}), and the orbital pseudospin is reduced to an Ising variable. At high temperature, the orbital is disordered, with each sites being equally populated by $d_{xz}$ and $d_{yz}$, as illustrated in Fig.~\ref{fig:oo_strong}(a). Consequently, the magnetic exchange is weak and there is no magnetic order. At low temperature, we have a ferro-orbital order of either $d_{xz}$ or $d_{yz}$ [Fig.~\ref{fig:oo_strong}(b) and (c)]. The effective spin model becomes an anisotropic $J_{1a}$-$J_{1b}$-$J_2$ Heisenberg model. If $J_2 > J_{1b}$, we have a stripe-like AFM order with the ordering wave vector $\bm{Q}=(\pi,0)$ or $(0,\pi)$.

We note that our analysis is based on strong anisotropy in the hopping amplitudes of the $d_{xz}$ and $d_{yz}$ orbitals. This assumption is confirmed by DFT calculations\cite{Lee2009} in which a similar scenario is also proposed. Experimentally inelastic neutron scattering results\cite{Zhao2009} give further support that the spin physics in iron pnictides is governed by a strongly anisotropic Heisenberg model. Now the question is whether orbital order can exist in the absence of the long-range magnetic order. First, we note that the orbital order is an Ising order whereas the spin order is an $O(3)$ order. In a true two-dimensional system, orbital order can occur at finite temperature whereas spin order only sets in at $T=0$. Turning on couplings along the third dimension, we can drive the magnetic transition close to the orbital transition. Nevertheless, there exists the nematic phase with nonzero orbital order but no magnetic order. This argument is also used by some theories on the basis of spin fluctuations.\cite{Fang2008,Xu2008} Numerically, such a sequence of phase transitions is found in both exact diagonalization\cite{Chen2009} and Monte-Carlo\cite{Applegate2012} studies of the spin-orbital model. Second, from our discussion, orbital order can be driven by other interactions, like electron-phonon and quadrupole-quadrupole interactions, in addition to the spin-orbital couplings. Therefore orbital order transition can have a distinct energy scale from the magnetic transition, and happen at a much higher temperature. This is indeed the case in many other transition metal oxides.\cite{Tokura2000}

\subsection{Orbital order from weak coupling}
\label{sec:weak}

\begin{figure}[t]
  \centering
  \includegraphics[width=\textwidth]{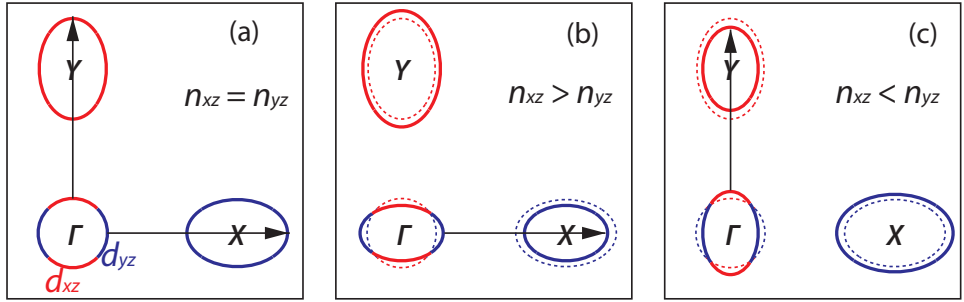}
  \caption{Illustration of the Fermi surface distortion from orbital order. (a) Fermi surfaces without orbital order. (b) Fermi surfaces with $d_{xz}$ ferro-orbital order. (c) Fermi surfaces with $d_{yz}$ ferro-orbital order. In (b) and (c) the original Fermi surfaces are shown by the dashed lines. The arrows represent the nesting instability between the hole and electron pockets. The color represents the dominant orbital at that location, with `red' corresponding to $d_{xz}$ and `blue' to $d_{yz}$.}
  \label{fig:oo_weak}
\end{figure}

In the previous subsection, we have described how orbital order arises in the strong-coupling limit. However, the normal state of the iron-based superconductors remains metallic.
As a consequence, the opposite weak-coupling limit is worthy of being studied as well.

Let us examine the Fermi surface of the iron-based superconductors. In the unfolded Brillouin zone that contains one Fe atom per unit cell, there is a circular hole pocket at the zone center $\Gamma = (0,0)$, whereas two elliptical electron pockets appear at $X=(\pi,0)$ and $Y=(0,\pi)$, respectively [see Fig.~\ref{fig:oo_weak}(a)]. Here for simplicity we have ignored the second hole pocket at $\Gamma$ and a third possible hole pocket at the zone corner $M=(\pi,\pi)$. In terms of a two-orbital model, the eigenstates on different locations of the Fermi surfaces
have different weights of $d_{xz}$ and $d_{yz}$ orbitals. One prominent feature here is that under a $\pi/2$ rotation, the shape of the Fermi surface is invariant, but the dominant orbital weight changes between $d_{xz}$ and $d_{yz}$. This condition is strictly enforced because the two orbitals exchange with each other under the $C_4$ rotation.

It is then obvious that an electron nematic order can be captured by a ferro-orbital order.\cite{Chen2010,Lv2011a} As shown in Fig.~\ref{fig:oo_weak}(b), if $d_{xz}$ orbital has a larger occupation number, the hole pocket will be elongated along the $x$ direction and shortened along the $y$ direction. Such a Fermi surface distortion is a
hallmark of the nematic order. Meanwhile, the electron pocket at $X$ shrinks while the one at $Y$ enlarges. Similarly for the $d_{yz}$ orbital order, the opposite takes place [Fig.~\ref{fig:oo_weak}(c)]. So in this multiorbital system, orbital order naturally leads to an electronic nematicity.

This ferro-orbital order has an important consequence in picking up the preferred wavevector of the stripe-like AFM order. In this itinerant model, the magnetic order is induced by the nesting between the hole and electron pockets. As we can see in Fig.~\ref{fig:oo_weak}(a), for the isotropic Fermi surface, the nesting can happen between either of the two electron pockets. Furthermore, due to the mismatch between the hole and electron pockets, the nesting condition is far from perfect and the nesting wavevector is usually incommensurate. In the presence of the ferro-orbital order,
the distorted Fermi surface enhances the nesting condition along one of the two directions with a commensurate nesting wavevector. In particular, $d_{xz}$ ferro-orbital order favors the $(\pi,0)$ magnetic order while the $(0,\pi)$ order arises in the $d_{yz}$ orbitally ordered state. This argument also gives a natural explanation why the structural phase transition always precedes the magnetic transition.

Very recently, Avci {\it et. al.} observed a novel first order transition below $T_N$ into a new magnetic order with $C_4$ symmetry restored 
in Ba$_{0.76}$Na$_{0.24}$Fe$_2$As$_2$.\cite{avci2013} While the exact configuration of this new magnetic state remains to be determined, it has been shown that a 
spin-nematic model could lead to several candidate states within a limited range of parameter space. 
We note that such 're-entry' of $C_4$ symmetry could still be consistent with an orbital model. Let us denote $M_1$ as the stripe-like AFM order with 
orbital order and $M_2$ as this new magnetic state with $C_4$ symmetry. Since $M_1$ and $M_2$ break different symmetries, a first order transition between them could naturally 
occur if they are competing orders. This is a general principle regardless the nature of the model. As a result, if a spin-nematic model could produce such a first order transition, 
so could an orbital model in principle. In fact, this again reflects that the mean-field phase diagram is insentitive to the details of the model
as long as the symmetries of the phases are captured correctly. Nevertheless, it is still intriguing to further explore the physical properties of this new magnetic state both 
experimentally and theoretically.

It is then clear that in order to resovle the microscopic mechanism driving the phase transitions in iron-based superconductors, 
the investigation on quantum fluctuations beyond the mean-field states is necessary. As will be shown in the next section,
the ferro-orbital order in a multiorbital system is in fact {\it identical} to the {\it Pomeranchuk instability} in the $d$-wave channel.\cite{oganesyan2001,pomeranchuk}
The elementary excitations in the proximity of the orbital ordering quantum critical point are unusual, leading to a non-Fermi liquid behavior which cannot be caused by 
the spin nematicity.

\section{Excitations near the orbital ordering quantum critical point}
\label{sec:excitation}
\subsection{Non-Fermi liquid behavior near quantum critical point in itinerant systems}

In this subsection, we briefly review the principle behind the non-Fermi liquid behavior emerging from the quantum criticality, which has been a hot research topic over the last decades.
Generally speaking, a phase transition can be described by a problem of minimization of the free energy $F=U-TS$, where $U$
is the internal energy of the system, $T$ the temperature, and $S$ is the entropy of the system.
In classical systems, the thermal fluctuations are dominant and hence, the phase transition is a compromise between states with lower $U$ and higher $S$.
At high temperature, $F$ is minimized by the state with higher entropy which has the most symmetries the system can have.
On the other hand, at low temperature a state with lower internal energy is favored.
In an interacting system, such a state could have less symmetries.
As a result, a phase transition occurs when the system spontaneously loses some symmetries at a
critical temperature $T_c$ as it is cooled down from high to low temperature.
An indicator for the phase transition is an {\it order parameter} which develops a non-zero expectation value only in the low temperature ordered state.

In real systems, we can have a tuning parameter, namely $g$, that we can use to change $T_c$. Typical examples for the tuning parameter are external pressure, magnetic field, and chemical dopings.
A quantum phase transition\cite{sachdev} is then referred to a transition occurring at a quantum critical point (QCP) defined at $T_c=0$ and $g=g_c$.
A surprising feature of this subject is that although the quantum phase transition is strictly at zero temperature, the associated quantum fluctuations near QCP
can fundamentally change the physical properties even at finite temperature.
This is why the field of quantum criticality has been so vibrant recently. Along with its theoretical novelty, it is also very relevant to the strongly correlated
materials.
A recent review on new progress in several QCPs of insulating phases can be found in Ref.~\refcite{xureview}.

While the quantum fluctuations associated with the QCP can be described by the dynamics of the order parameters, the situation is further complicated in an itinerant system.
In an insulating phase, by definition, all the quasiparticle spectrum is gapped, and as a consequence, the low energy physics contains only the collective excitations
from the order parameter dynamics.
If the QCP, however, resides in an itinerant system, the existence of the gapless quasiparticles can affect the dynamics of the order parameter and in feedback, the quasiparticle properties can also be altered fundamentally.
The pioneering work in demonstrating this idea was done by Hertz\cite{hertz1976} and later improved upon by Millis\cite{millis1993}.

Consider a Fermi liquid near a ferromagnetic QCP. Since the ferromagnetic order is at zero momentum ($\vec{q}=0$), we analyze the low energy excitations near $\vec{q}=0$.
The particle-hole continuum from quasiparticles is bounded below $Max \big[\epsilon(\vec{k}+\vec{q}) - \epsilon(\vec{k})\big] \sim \vert \vec{q}\vert$,
where $\epsilon(\vec{k})$ is the quasiparticle energy at momentum $\vec{k}$.
Given that the dispersion of the magnon in a ferromagnet goes like $q^2$, as the ferromagnetic QCP is approached from the Fermi liquid side, a branch of spin collective modes
emerges with a dispersion of $\delta(g) + a q^2$, where $\delta(g)$ can be viewed as the mass of the magnon that vanishes at the
QCP ($\delta(g\to g_c)\to 0$).
Clearly, right at the QCP the whole branch of the spin excitation modes lies completely inside the
particle-hole continuum and hence, these collective modes can no longer have infinite life time and will eventually decay into uncorrelated particle-hole pairs.
As shown by Hertz and Millis, this {\it overdamping} effect brings in a correction to the {\it dynamical exponent} $z$ and can be captured even at the
random-phase approximation (RPA) level.

Now we are ready to study the fate of the quasiparticles. Since these overdamped spin modes near zero momentum still have non-zero spectral weights,
quasiparticles can be scattered off them.
As a result, the life time of the quasiparticles, which is inversely proportional to the imaginary part of the quasiparticle self-energy ${\rm Im}\Sigma(\omega)$,
is no longer simply determined by the phase space argument as for a Fermi liquid\cite{mermin}.
Rather, it is mostly determined by the dimension of the system as well as $z$ of the overdamped collective modes, ${\rm Im} \Sigma(\omega) \sim \omega^{d/z}$, and in the
current two-dimensional ferromagnetic QCP case, ${\rm Im} \Sigma(\omega) \sim \omega^{2/3}$.
This is the milestone result signaling the violation of the Fermi liquid behavior (${\rm Im} \Sigma(\omega) \sim \omega^2$) induced by the quantum criticality.

Although the example given above is for a ferromagnetic QCP, the same theory can be applied to all the zero-momentum ordered states transitioned into from a Fermi liquid since they fall into the same effective field theory in Landau-Ginzburg formalism.
Another example is the two-dimensional nematic quantum fluid first proposed by Oganesyan {\it et al.},\cite{oganesyan2001} which motivated a number of theoretical efforts.\cite{metzner2003,yamase2004,lawler2006,kao2007,yamase2012}
As discussed in Sec.~\ref {sec:orbital}, the nematic order is a translationally invariant phase breaking the rotational symmetry in which the Fermi surface is spontaneously distorted along one particular direction in the two-dimensional plane.
In fact, the instability toward the change of the Fermi surface was studied a long time ago in 1958.
Pomeranchuk\cite{pomeranchuk} first derived the criteria for the Fermi surface instabilities induced by the interactions at higher angular momentum channels described by the Landau parameters $F_n$. The nematic order is one of these Pomeranchuk instabilities which is in the $d$-wave interaction channel $F_2$, and it has received a lot of attention in the last decade due to the discoveries of anisotropic phases in several systems.\cite{fradkin2010,lilly1999,du1999,kivelson2003}
What was unexpected back at the time of Pomeranchuk is that the nematic quantum fluid could exhibit a non-Fermi liquid behavior in the same way as the
itinerant ferromagnet, except that the interaction in this case would be from the charge instead of the spin.
The reasoning is as follows. Because the nematic order still preserves the translational invariance, it must be an order at zero momentum.
Moreover, since it breaks a continuous symmetry of rotation, it will develop Goldstone modes with a dispersion of $q^2$.
Finally, since the Fermi surface is only distorted but not killed, there are still itinerant electrons and consequently the particle-hole continuum is still present.
As a result, all the physics discussed in the ferromagnetic QCP can be applied to the nematic QCP despite some differences in the details.
Indeed, the overdamped collective modes with $z=3$ as well as the non-Fermi liquid behavior were found by Oganesyan {\it et al.}\cite{oganesyan2001} using RPA theory, and later Lawler {\it et al.} confirmed these findings using higher dimensional bosonization.\cite{lawler2006}
We emphasize that the consistency between the results from RPA and higher dimensional bosonization is remarkable since the former is perturbative while the latter is non-perturbative. This suggests that RPA theory is asymptotically exact at long wavelength, and an analysis based on RPA theory for the quantum criticality near zero momentum is a reasonable starting point despite its perturbative nature.

There is a concern that since the materials of interest are crystalline and do not have continuous rotational symmetry, the Goldstone modes are not massless, which would be at odds with the occurrence of the non-Fermi liquid. The main effect of this is that at zero temperature,
in a lattice model the non-Fermi liquid behavior goes away immediately in the nematic ordered phase, while in a continuous model it survives throughout the ordered phase. However, as long as the transition is of a second order, the critical fluctuations are still overdamped and massless at the QCP in a lattice model, and the non-Fermi liquid behavior still prevails. At finite temperature, if the thermal fluctuations are large enough, i.e., $k_B T_c > m$ where $m$ is the mass of the Goldstone mode, the non-Fermi liquid behavior still emerges.
The non-Fermi liquid behavior near the nematic QCP at zero and finite temperature in a lattice model has been studied.\cite{kao2007,yamase2012}

\subsection{Analysis on a two-orbital model}
As discussed in the previous sub-section, the finger print of the non-Fermi liquid behavior is the collective mode arising from critical fluctuations near the QCP with a dynamical exponent of $z=3$. This is commonly the case for a zero-momentum ordered state.
In this section, we will review the study on a two-orbital model which is the minimal model for the iron pnictides and bilayer Sr$_3$Ru$_2$O$_7$.
The key point here is that the orbital order in a two-orbital model is {\it identical} to the nematic order in a single-band lattice model.
Hence they share the same physics, including the non-Fermi liquid behavior near the QCP.

To see this connection, we start from the Fermiology of the two-orbital model discussed in Sec. 2. Within the framework of the Fermi liquid theory,
the interactions that are relevant should be the ones between states on the Fermi surface. Without a loss of generality,
the eigenstate wavefunction on the $\alpha$ Fermi surface can be generally written as:\cite{leewc2009sr327}
\be
|\psi_{\alpha,\sigma} (\vec p)\rangle = \big(\cos \phi_{\vec p} |d_{xz}(\vec p)\rangle +\sin \phi_{\vec p} |d_{yz}(\vec p)\rangle\big)
\otimes \chi_\sigma
\ee
where $\phi_{\vec p}$ is the hybridization angle between $d_{xz}(\vec p)$ and $d_{yz}(\vec p)$ at $\vec p$ and $\chi_\sigma$ is spin eigenstate.
In real space, the $C_4$ symmetry is enforced by the interchange of $d_{xz}$ and $d_{yz}$ orbitals up to a sign under a $\frac{\pi}{2}$ rotation,
$\hat{{\rm R}}_{\frac{\pi}{2}}$.
This places a relation between the eigenstate wavefunctions at $\vec p$ and $\vec p^\prime=\hat{{\rm R}}_{\frac{\pi}{2}} \vec p$ of
$\phi_{\vec p^\prime} = \phi_{\vec p} + \frac{\pi}{2}$.
Following the standard procedure, we evaluate the interaction $V(\vec p_1,\vec p_2)$ between states at $\vec p_1$ and $\vec p_2$ on the Fermi surface as
\be
\langle \psi_{\alpha,\sigma} (\vec p_1), \psi_{\alpha,\sigma} (\vec p_2)| V(\vec p_1,\vec p_2) | \psi_{\alpha,\sigma} (\vec p_2),\psi_{\alpha,\sigma} (\vec p_1)\rangle.
\label{interaction}
\ee
Eq.~(\ref{interaction}) can be evaluated by the Hartree-Fock approximation which gives
\bea
V_\mathrm{Hartree} &=& V(\vec p_1,\vec p_2),\nn\\
V_\mathrm{Fock} &=& -V(\vec p_1,\vec p_2) \vert \langle \psi_{\alpha,\sigma} (\vec p_1)|\psi_{\alpha,\sigma} (\vec p_2)\rangle\vert^2.
\eea
Clearly, the Hartree term is not affected by the multiorbital nature, but the effect of the orbital hybridization comes in the
Fock exchange part, leading to the effective Landau interaction function,
\bea
f^s(\vec p_1,\vec p_2)&=& V(\vec q=0)
-\frac{1}{4}[1+\cos 2
(\phi_{\vec p_1}-\phi_{\vec p_2})]
\times V(\vec p_1,\vec p_2),
\nn \\
f^a(\vec p_1, \vec p_2)&=& -\frac{1}{4}[1+\cos 2
(\phi_{\vec p_1}-\phi_{\vec p_2})] \times V(\vec p_1,\vec p_2).
\eea
Now it becomes apparent that because of the form factor $\cos 2 (\phi_{\vec p_1}-\phi_{\vec p_2})$,
the orbital hybridization induces a $d$-wave Landau interaction {\it regardless the original form of} $V(\vec p_1,\vec p_2)$.
In other words, the tendency toward the nematic order, or equivalently an orbital order, is naturally enhanced in such a two-orbital system.

Applying the same procedure outlined above, one can readily show that the two-orbital model can be mapped to a two-band model with a large $F_2$ interaction on its Fermi surfaces. Such a mapping has to be true because orbital degrees of freedom are only well-defined on a lattice.
As a result, it will eventually evolve into a band model when approaching the continuum limit. Since there is no phase transition induced simply by taking the continuum limit, the orbital model has the same physics as its corresponding band model. 
This indicates that the orbital order in a weak-coupling itinerant system is {\it identical} to the nematic order from the $d$-wave Pomeranchuk instability. 
This is different from the strong coupling limit in which the orbital order is an Ising-like order as discussed in Sec. \ref{sec:strong}. 
This should not be surprising since quite generally the hybridizations between different orbitals are in the single particle part of the Hamiltonian, not in the interaction part. 
As a result, the orbital index is not a good `quantum number' in the weak coupling limit but remains approximately a good one in the strong coupling limit.

There have been several studies on the two-orbital model beyond the mean-field theory.
A renormalization group analysis showed that the orbital order instability can be largely enhanced by spin fluctuations, which has been exploited to explain the
magnetic field-induced nematicity observed in Sr$_3$Ru$_2$O$_7$.\cite{ohno2013,tsuchiizu2012}
Lo {\it et al.}\cite{lo2012} have analyzed the two-orbital model using higher dimensional bosonization.
Near the orbital ordering QCP, it is found that a branch of overdamped collective modes with $z=3$ does emerge. This provides a strong foundation for the occurrence of the non-Fermi liquid behavior in the iron pnictides which will be reviewed in the next section.

\subsection{RPA study on a five-orbital model}
Iron pnictides are known for their complicated multiorbital structure. It has been suggested that in the unfolded Brillouin zone, a model containing
five orbitals is a much better choice to capture the details in the electronic structure.\cite{graser2009}
Although the orbital order involves mostly the $d_{xz}$ and $d_{yz}$ orbitals, the couplings to other orbitals can substantially modify the hybridization between $d_{xz}$ and $d_{yz}$. Therefore a theory based on a five-orbital model capturing enough detail is certainly necessary to demonstrate the non-Fermi liquid behavior. Fortunately, since the RPA theory is a reliable starting point as explained in Sec.~3.1, one can directly calculate the single particle self energy\cite{leewc2012nfl} at the RPA level with a multiorbital Hubbard model.\cite{graser2009}

\begin{center}
\begin{figure}[t]
\includegraphics{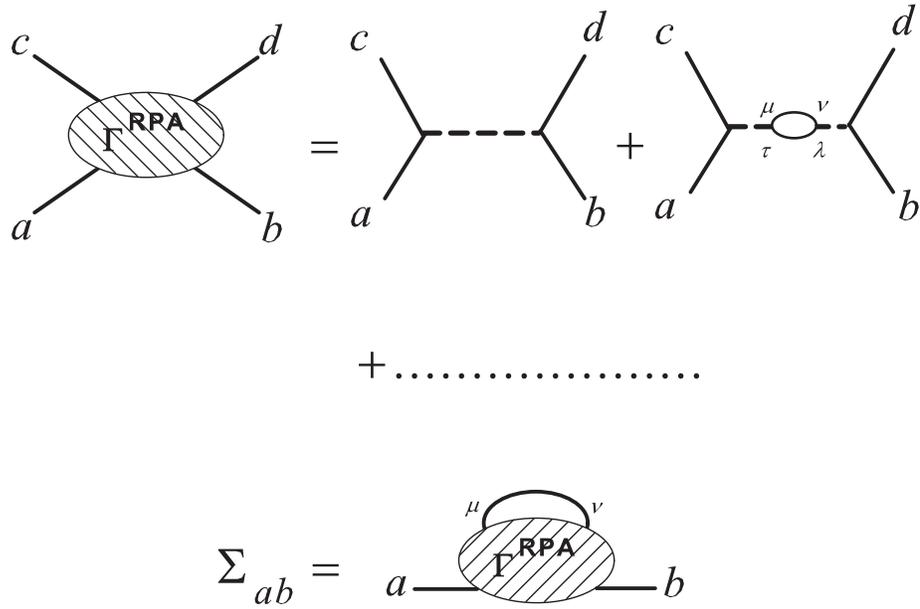}
\caption{\label{fig:feynman} Feynman diagrams for the single particle self energy at the RPA level. The dashed line refers to the interactions between different orbitals,
and the solid line is the single particle propagator $G_{ab}$. All the symbols refer to an index of orbital and spin. The Greek symbols are
internal indices which are summed over all orbitals and spins.}
\end{figure}
\end{center}

The computational scheme for the self energy $\Sigma_{ab}(\vec{k},\omega)$ can be depicted by the Feynman diagram plotted in Fig.~\ref{fig:feynman},
and we refer the readers to Ref.~\refcite{leewc2012nfl} for details. It is shown\cite{leewc2012nfl} that ${\rm Im} \Sigma_{ab}(\vec{k}_F,\omega)$ changes from a Fermi liquid ($\sim \omega^2$) to a non-Fermi liquid ($\sim \omega^\alpha, \alpha < 1$) behavior as the orbital ordering quantum critical point is approached within a realistic range of interaction parameters for the multiorbital Hubbard model. This further confirms that the important physics from the two-orbital model survives in a complicated five-orbital model. In Sec.~\ref{sec:expt}, we will discuss that such a non-Fermi liquid behavior gives a natural explanation to the novel enhancement of the zero bias conductance observed in a recent point contact measurement.\cite{arham2012}

It should be noted that in this five-orbital Hubbard model, contributions to the single particle self energy from the spin interaction are actually not small. As a result, it is an open debate as to whether the orbital ordering is caused from charge or spin interactions, and with the current model, there seems to be no hope in resolving this issue. Nevertheless, if one includes the nearest-neighbor interaction, the contributions from charge interaction can be enhanced. This is demonstrated\cite{puetter2010} within the context of Sr$_3$Ru$_2$O$_7$ which is another intriguing material exhibiting the anisotropic phase under magnetic field. Moreover, Kontani {\it et al.} showed that including the
electron-phonon coupling as well as the vertex corrections beyond the RPA level can boost up the orbital order instabilities in the charge channel which may lead to an even richer phase diagram.\cite{kontani1,kontani2,kontani3}.
Since these new additions to the multiorbital Hubbard model are all realistic and not negligible in general, the orbital ordering from charge interaction should be more important than it now appears, at least for some families of the iron pnictides.

The non-Fermi liquid behavior could be a strong indicator for the orbital fluctuations because fluctuations from spin nematicity do not seem to
produce the same result. The reason is given as follows.
As described in Sec.~\ref{sec:orbital},
in the spin nematic scenario the nematic phase breaks the $Z_2$ symmetry arising from the degeneracy of two possible ordering wavevectors $(\pi,0)$ and $(0,\pi)$.
Following this theory, the critical fluctuations near the nematic QCP would mostly arise from the spin fluctuations around finite momenta $(\pi,0)$ and $(0,\pi)$, which, according to Hertz-Millis theory, usually have a much weaker capability of inducing non-Fermi liquid behavior. To date, we are not aware of any theoretical study proposing a non-Fermi liquid behavior arising from spin nematicity.

\section{Experimental signatures for orbital fluctuations}
\label{sec:expt}

\subsection{Point contact spectroscopy}

\subsubsection{Introduction}

A point contact is simply a contact between two metals whose characteristic size $d$ is much less than the electron elastic and inelastic mean free paths: $l_{el},l_{in} \gg d$. Point contact spectroscopy (PCS) studies the non-linearities of the current-voltage (I-V) characteristics of these metallic constrictions. The bias voltage applied to the point contact junction determines the energy scale of the scattering process. If both metals are ohmic, the I-V curve is linear and the differential conductance $dI/dV$ is constant. However, when scattering processes are present, they show up in the PCS spectrum. PCS was initially used to study electron-phonon interactions\cite{Yanson} in metals. If one of the metals is replaced by a superconductor, making a N-S junction, Andreev reflection\cite{Andreev} is observed. In recent years, PCS has also been used to probe heavy fermion compounds where it is sensitive to the onset of the Kondo lattice that appears as a Fano line shape\cite{Park2} and to the hybridization gap in the heavy fermion $\rm{URu_2Si_2}$.\cite{Park3} PCS has also been used extensively to study the iron based superconductors and provides evidence for orbital fluctuations in the normal sate of underdoped $\rm{Ba(Fe_{1-x}Co_x)_2As_2}$, parent $\rm{SrFe_2As_2}$ and $\rm{Fe_{1+y}Te}$.\cite{arham2012}

\subsubsection{Conductance Spectra}

\begin{figure}[t]
	\includegraphics{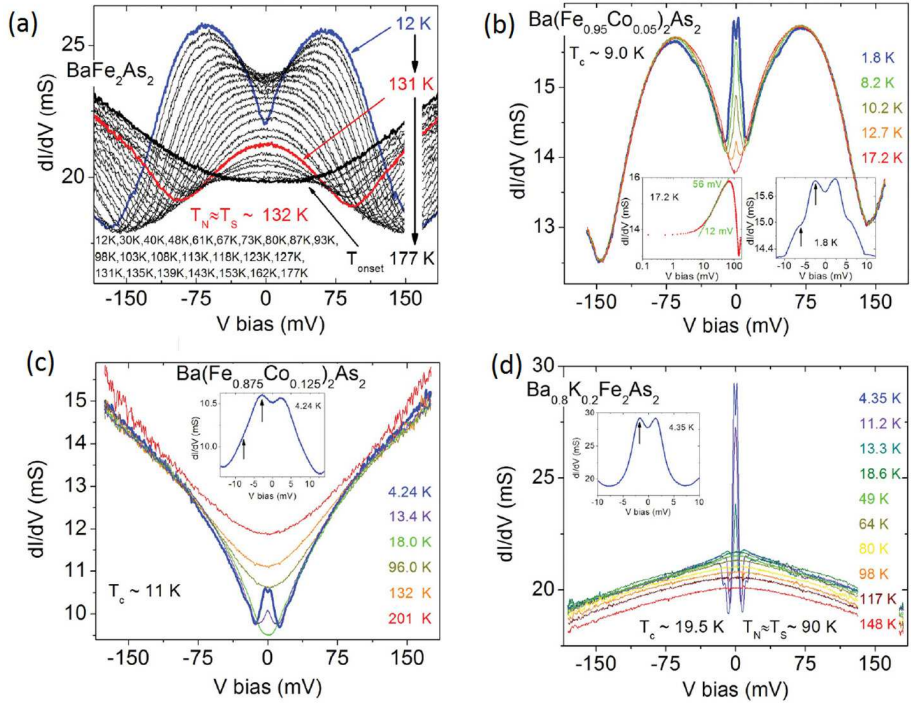}
	\caption{From Ref.~\protect\refcite{arham2012}, copyright 2012 by the American Physical Society. (a) Conductance spectra for $\rm{BaFe_2As_2}$. Conductance enhancement with peaks at $\sim$ 65 mV superimposed on a parabolic background was observed at low temperatures. The peaks moved in as the temperature was increased and the enhancement survived well above $T_S$ (red curve). (b) $\rm{Ba(Fe_{0.95}Co_{0.05})_2As_2}$ displays a coexistence of magnetism and superconductivity. At low temperatures, clear Andreev peaks were observed [right inset (b); the arrows are pointing out the Andreev peaks]. A conductance enhancement with peaks at $\sim$ 65 mV coexisted with the Andreev spectra and evolved with temperature as it did for $\rm{BaFe_2As_2}$. This enhancement increased logarithmically near zero bias [left inset (b)]. The overdoped compound $\rm{Ba(Fe_{0.875}Co_{0.125})_2As_2}$ showed Andreev spectra below $T_c$. It did not have conductance peaks at higher bias values like the Co underdoped compounds. (d) The hole underdoped $\rm{Ba_{0.8}K_{0.2}Fe_2As_2}$ has a coexistence of superconductivity and magnetism. It showed Andreev spectra below $T_c$ and no higher bias conductance enhancement. This was in contrast to the data obtained from electron underdoped $\rm{Ba(Fe_{1-x}Co_x)_2As_2}$ [Fig.~\ref{fig:pcs}(b)].}
\label{fig:pcs}
\end{figure}

Arham {\it et al.}\cite{arham2012} reported conductance spectra on $\rm{Ba(Fe_{1-x}Co_x)_2As_2}$, spanning the entire phase diagram. Part of their data is reproduced in Fig.~\ref{fig:pcs}. For the undoped parent compound [Fig.~\ref{fig:pcs}(a)], at the lowest temperature (blue curve), they saw a dip at zero bias and two asymmetric conductance peaks at $\sim$ 65 mV. This double peak feature was superimposed on a parabolic background. As the temperature was increased, the dip at zero bias filled up, the conductance peaks moved inward, and the bias voltage range of the conductance enhancement decreased. No dramatic change occurred as the structural transition temperature, $T_S$, was crossed (red curve). The enhancement eventually disappeared at 177 K, more than 40 K above $T_S$.

For underdoped $\rm{Ba(Fe_{1-x}Co_x)_2As_2}$ [Fig.~\ref{fig:pcs}(b)], where superconductivity coexists with long-range magnetic order, they observed Andreev reflection at low voltage biases in the superconducting state. However, just like the parent compound, two conductance peaks occurred at $\sim$ 65 mV. Above the onset temperature of the superconducting transition $T_c$, Andreev reflection completely disappeared and the high bias conductance evolved just as it did for $\rm{BaFe_2As_2}$. The right inset in Fig.~\ref{fig:pcs}(b) shows a zoom in of the Andreev reflection features while the left inset plots the conductance spectra above $T_c$ on a log plot.

For overdoped $\rm{Ba(Fe_{1-x}Co_x)_2As_2}$ [Fig.~\ref{fig:pcs}(c)], Andreev reflection was observed in the superconducting state, but unlike the underdoped compounds, no higher bias conductance peaks were detected. Above $T_c$, a parabolic background remained which flattened with further increase in temperature.

Conductance spectra were also presented on hole underdoped $\rm{Ba_{1-x}K_{x}Fe_2As_2}$ [Fig.~\ref{fig:pcs}(d)]. The sample has a coexistence of magnetism and superconductivity ($T_N$ = $T_S$ $\sim$ 90 K, $T_{c}$ $\sim$ 20 K). Below $T_c$ clear Andreev reflection was observed. Above $T_c$, Andreev reflection disappeared, leaving a downward facing background that did not change with any further increase in temperature. This is remarkably different from the situation in electron underdoped $\rm{Ba(Fe_{1-x}Co_x)_2As_2}$.

As comparisons, $\rm{CaFe_2As_2}$ and $\rm{SrFe_2As_2}$ were also probed. The data for $\rm{SrFe_2As_2}$ [Fig.~2(a) in Ref.~\refcite{arham2012}] were very similar to those of $\rm{BaFe_2As_2}$, with conductance enhancement around zero bias that sets in before the magnetostructural transition. However, for $\rm{CaFe_2As_2}$ [Fig.~2(b) in Ref.~\refcite{arham2012}], the conductance enhancement disappeared around 100-110 K, which is much lower than the magnetostructural transition temperature, 170 K.

Similar features were also observed in the parent compound for the iron chalcogenide superconductor. $\rm{Fe_{1.13}Te}$ showed a conductance enhancement that survived above the magnetic and structural transition temperatures [Fig.~2(c) in Ref.~\refcite{arham2012}]. The conductance enhancement was observed till 75 K ($T_N$ = $T_S$ $\sim$ 59 K).

To summarize the work of Arham {\it et al.},\cite{arham2012} $\rm{BaFe_2As_2}$, $\rm{SrFe_2As_2}$, underdoped $\rm{Ba(Fe_{1-x}Co_x)_2As_2}$ and $\rm{Fe_{1+y}Te}$ exhibited a conductance enhancement that sets in above $T_S$, $\rm{CaFe_2As_2}$ only showed the enhancement below $T_S$ while $\rm{Ba_{0.8}K_{0.2}Fe_2As_2}$ did not show any conductance enhancement. Overdoped $\rm{Ba(Fe_{1-x}Co_x)_2As_2}$ does not have a $T_S$ and only showed Andreev spectra below $T_c$. The high bias background for all compounds except for $\rm{Ba_{0.8}K_{0.2}Fe_2As_2}$ was an upward facing parabola.

\subsubsection{Discussion}

Arham {\it et al.}\cite{arham2012} noticed a correlation between the presence of conductance enhancement around zero bias and in-plane resistivity anisotropy in the compounds. For detwinned underdoped $\rm{AEFe_2As_2}$ it has been shown that below $T_S$ a resistivity anisotropy exists. \cite{Chu2010,Tanatar,Fisher,Blomberg} Above $T_S$ there is notable anisotropy for AE = Ba, negligible anisotropy for AE = Sr, and no anisotropy for AE = Ca. Detwinned $\rm{Fe_{1+y}Te}$ also shows a resistivity anisotropy above the structural transition.\cite{Jiang} The anisotropy above $T_S$ is sensitive to the uniaxial force required to detwin the samples. Detwinned underdoped $\rm{Ba_{1-x}K_{x}Fe_2As_2}$ does not show any anisotropy at all, either below or above $T_S$.\cite{Ying}

\begin{figure}[t]
		\includegraphics{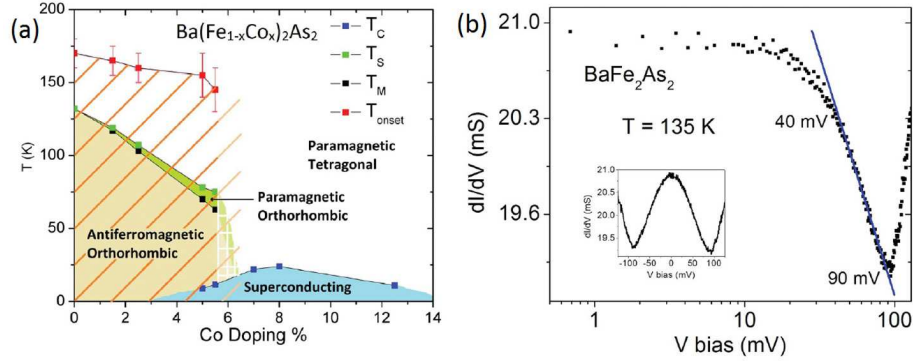}
	\caption{From Ref.~\protect\refcite{arham2012}, copyright 2012 by the American Physical Society. (a) Phase diagram for $\rm{Ba(Fe_{1-x}Co_x)_2As_2}$ marking a new line on the underdoped side showing the temperature below which the conductance enhancement was observed by PCS. (b) Conductance above $T_S$ for $\rm{BaFe_2As_2}$ followed a log dependence from $\sim$ 40 mV to $\sim$ 90 mV.}
	\label{fig:pcs-phase}
\end{figure}

The presence or absence of the in-plane resistivity anisotropy matches whether or not a conductance enhancement is detected. The correlation of the conductance enhancement with the resistivity anisotropy indicates they are likely caused by the same underlying physics. Arham {\it et al.}\cite{arham2012} constructed a revised phase diagram for $\rm{Ba(Fe_{1-x}Co_x)_2As_2}$ marking a new line on the underdoped side showing the temperature below which the conductance enhancement was observed [Fig.~\ref{fig:pcs-phase}(a)].

Theoretical work by Lee {\it et al.}\cite{leewc2012nfl} (reviewed in Sec.~3) showed that orbital fluctuations above $T_S$ were expected to provide extra contributions to the single-particle density of states (DOS) at zero energy.\cite{lawler2006} The DOS followed a log dependence as the energy was increased. Arham {\it et al.} compared their data with this prediction and found that the conductance enhancement for $\rm{BaFe_2As_2}$ above $T_S$ followed a log dependence from $\sim$ 40 mV to $\sim$ 90 mV [Fig.~\ref{fig:pcs-phase}(b)]. They obtained similar fits above $T_S$ for $\rm{SrFe_2As_2}$ and $\rm{Fe_{1.13}Te}$. Furthermore, the absence of similar effects in the data on $\rm{Ba_{0.8}K_{0.2}Fe_2As_2}$ was consistent with the prediction that crystals that did not show the resistivity anisotropy would also not exhibit the excess conductance due to orbital fluctuations. Their data therefore strongly indicated that the enhancement in conductance observed was a consequence of orbital fluctuations.

It should be kept in mind that the conductance ($dI/dV$) measured by point-contact spectroscopy does not directly correspond to the density of states. PCS data is a convolution of the Fermi velocity and the energy-dependent density of states along with any scattering processes that might be present. For normal metals, the Fermi velocity and the density of states are inversely related and cancel each other out.\cite{Harrison} There is a lack of theoretical models for interpreting PCS data on correlated metals, where the DOS are energy dependent and do not cancel out with the Fermi velocity when $dI/dV$ is measured. A theory considering both the energy dependence of the electronic DOS and scattering processes would be extremely helpful in obtaining a better understanding of the experimental data.

\subsection{Magnetic torque measurement on $\rm{BaFe_2(As_{1-x}P_x)_2}$}

\subsubsection{Introduction}

Torque magnetometry is an excellent method to detect the presence of $C_4$ symmetry breaking novel electron matter in a tetragonal crystal lattice.
The quantity to be measured is the torque defined as $\vec{\tau}=\mu_0 V \vec{M} \times \vec{H}$, where $\mu_0$ is the permeability of vacuum, $V$ the crystal volume,
and $\vec{M}$ is the magnetization induced by magnetic field $\vec{H}$.
When $\vec{H}$ is rotated within the tetragonal $a$-$b$ plane, it can be shown\cite{Matsuda} that the torque has a component of
$\tau_{2\phi}=\frac{1}{2}\mu_0H^2V[(\chi_{aa}-\chi_{bb})\sin2\phi-2\chi_{ab}\cos2\phi]$, where $\phi$ is the azimuthal
angle measured from the $a$ axis and $\chi_{ij}$ is determined by $M_i=\sum_j{\chi_{ij}H_j}$. When the system has tetragonal symmetry, $\chi_{aa}=\chi_{bb}$ and $\chi_{ab}=0$, making $\tau_{2\phi}=0$.
A non-zero $\tau_{2\phi}$, meaning either $\chi_{aa}\neq\chi_{bb}$ or $\chi_{ab}\neq0$, signals tetragonal ($C_4$) symmetry breaking.

\subsubsection{Experimental data and discussion}

Kasahara {\it et al.}\cite{Matsuda} carried out torque magnetometry measurements on BaFe$_2$(As$_{1-x}$P$_x$)$_2$ spanning the entire phase diagram.
They observed that $\tau_{2\phi}$ obtained a non-zero value at an onset temperature $T^*$ higher than $T_S$, indicating a nematic phase transition occurring above the structural phase
transition. They further showed that the onset temperature, $T^*$, of this new nematic phase was suppressed with increasing P doping but persisted well into the overdoped side.
Above $T^*$ the torque signal is isotropic.

The fact that $\tau_{2\phi}$ becomes non-zero before the structural transition temperature is a clear indication of the presence of an electron nematic phase. Kasahara {\it et al.}\cite{Matsuda} noticed that below $T^*$, the function form of the $\tau_{2\phi}$ was: $\tau_{2\phi}= A_{2\phi}\cos2\phi$. This meant that $\chi_{aa}=\chi_{bb}$, $\chi_{ab}\neq0$, and the nematicity was along the [110] (Fe-Fe bond) direction.

Kasahara {\it et al.} also analyzed x-ray data on these crystals using a two-peak fitting procedure (Fig.~3 in Ref.~\refcite{Matsuda}).
They observed a very small but finite lattice distortion in the temperature range $T_S<T<T^*$. They argued that this was due to a linear coupling between the order parameter governing the orthorhombic lattice distortion and the one representing electronic nematicity. They postulate that the true thermodynamic transition thus occurs at $T^*$ and the one at $T_S$ is a meta-nematic transition.

The origin of the electronic nematicity occurring at $T^*$ is still under debate.
The magnetic field applied during the torque magnetometry measurement breaks the $C_4$ symmetry itself since it points along one direction on the two-dimensional plane. Thus there is an uncertainty whether the transition occurring at $T^*$ is a true second order phase transition or not.
Nevertheless, there is no doubt that certain fluctuations associated with the nematicity must be present above the structural phase transition.
While spin fluctuation is a plausible candidate for causing nematicity in the underdoped samples, the observation of nematicity in the region far from the antiferromagnetic phase casts a strong doubt on this. Keeping this in mind, orbital fluctuations might be a better candidate for causing the nematic phase. In particular, it is intriguing to see that $T^*$ for the parent compound BaFe$_2$As$_2$ is remarkably consistent with the onset temperature for the enhancement of the zero-bias conductance discussed in Sec.~4.1. As a result, in the orbital-based scenario, $T^*$ would be the temperature for a crossover at which the orbital fluctuations become important, instead of the transition temperature for a second order phase transition.

\subsection{Inelastic neutron scattering measurement on $\rm{Fe_{1+y-x}(Ni/Cu)_xTe_{0.5}Se_{0.5}}$}

\subsubsection{Introduction}
Inelastic neutron scattering measurement on unconventional superconductors provides information on their bosonic (phonons and magnons) excitation spectra. After intensive international efforts in applying this technique to cuprates, heavy fermions, and iron based superconductors,
the behavior of the spin excitations for magnetic fluctuation driven superconductors has been well established.
In the normal state, the low energy spin fluctuations usually have a peak feature near a certain wavevector which is typically the ordering wavevector
in the parent compound, e.g., $(\pi,\pi)$ for cuprates and $(\pi,0)$ for the iron based superconductors.
The appearance of this peak is an indicator of the strong spin fluctuations inherited from the magnetic order in the parent compounds.
On cooling below $T_c$, the low energy spin fluctuations at the same wavevector develop a gap due to the superconductivity, and the spectral weight gets shifted to a pronounced resonance peak.
However, qualitatively, the magnetic spectrum above and below $T_c$ tends to be the same.
Moreover, if there is no other phase transition in the normal state, the magnetic spectrum is unchanged with rising temperature, and the only prominent effect is the broadening of the spectral weights due to the thermal fluctuations.
A nice review can be found in Ref.~\refcite{scalapino2012}.

\subsubsection{Experimental Data}

Xu {\it et al.}\cite{Tranquada} probed the low energy spin fluctuations in $\rm{Fe_{1+y-x}(Ni/Cu)_xTe_{0.5}Se_{0.5}}$ using inelastic neutron scattering. The Ni/Cu substitution of Fe was done to perturb the system and reduce $T_c$. The unperturbed system has a $T_c$ of 14 K, 2$\%$ Ni brings it to 12 K, 4$\%$ Ni to 8 K, while 10$\%$ Cu suppresses superconductivity entirely. The magnetic neutron scattering intensity was probed near the $\bm{M}$ point, $\bm{Q}_{AF}=(0.5,0.5,0)$, along with its
evolution along the wave vector $\bm{q}=[1,-1,0]$. It was found that the temperature dependence of the magnetic excitation spectrum was qualitatively different
for superconducting and non-superconducting samples.
In the superconducting sample ($4\%$ Ni doping), the low energy spin excitation below $T_c$ showed a commensurate U-shaped dispersion, with the bottom of the U close to the spin resonance energy mode. As expected, the only difference on crossing $T_c$ was the change in the spectral weight of the resonance peak. However, at 100 K, the spectra looked remarkably different, showing two columns centered around incommensurate wavevectors
$\bm{Q}=(0.5\pm\delta,0.5\mp\delta,0)$ at low energies. Xu {\it et al.} described it as looking like the `legs of a pair of trousers'.
For the non-superconducting sample ($10\%$ Cu doping), the trousers shaped spectra were obtained at both low and high temperatures.
In other words, the prerequisite condition for superconductivity in this family is the occurrence of an incommensurate-to-commensurate transformation in the magnetic excitation spectrum in the normal state. For the superconducting samples, this crossover appeared to occur close to 3$T_c$.

\subsubsection{Discussion}

The question confronting Xu {\it et al.} was how the low energy magnetic spectra could evolve from two incommensurate vertical columns at $T \gg T_c$ to a commensurate U-shaped dispersion at low temperatures. If one considers only the spin interactions, such a transformation seems to be impossible without a magnetic transition. However, it can be naturally understood if there exists another energy scale which is distinct from the spin sector but could affect the spin excitation. As shown in Sec. 2, in an itinerant system orbital order can change the Fermi surface. Therefore the spin excitation spectrum, which is basically the flip of the electron spin as it is scattered between states on different parts of the Fermi surface, can be changed qualitatively by the orbital order and related fluctuations. This has been demonstrated by Lee {\it et al.},\cite{Tranquada2} and the incommensurate-to-commensurate transformation is reasonably reproduced from a five-orbital model including Gaussian fluctuations to account for the low energy fluctuating orbital correlations. This orbital correlation-induced transformation can be viewed as an analogy of the stripe-induced neutron anomaly in cuprates, in which the stripe order or correlation renders the neutron peak from $(\pi,\pi)$ to incommensurate wavevectors.

\section{Concluding Remarks}
In this paper, we have reviewed some topics of the orbital physics originating from degenerate $d_{xz}$ and $d_{yz}$ orbitals.
Although the current focus is on understanding some non-trivial experimental results in the iron pnictides, these theories are generally applicable to all systems having degenerate $d_{xz}$ and $d_{yz}$ orbitals on the Fermi surfaces, e.g., Sr$_3$Ru$_2$O$_7$.
While we fully recognize the importance of spin physics in the iron pnictides, our purpose is to point out that there exist some features and experimental data which seem to be understood more naturally by the orbital physics.
In particular, since iron pnictides are generally less correlated compared to cuprates, no energy scale is really dominant over any other.
This strongly suggests that many physical properties are probably not universal and as a consequence,
it is highly possible that some families of the iron pnictides have strong orbital physics character while others do not, depending on their detailed electronic structures.
To achieve a better understanding, more spectroscopic measurements are certainly necessary. For example,
a recent Raman measurement has shown a superconductivity-induced peak in the B$_{1g}$ channel in Ba$_{0.6}$K$_{0.4}$Fe$_2$As$_2$, which was originally interpreted as the existence of a competing $d$-wave superconducting component.\cite{kretzschmar2012}
One of us, however, showed that the orbital fluctuations in the superconducting state could lead to such a peak as well.\cite{leewc2013}
Nevertheless, it appears to be surprising that superconductivity is ubiquitous among the iron pnictides despite this non-universality. If it turns out that superconductivity does arise from a normal state with strong orbital fluctuations in some families of the iron based superconductors, a unified picture for the pairing mechanism will never be complete without the orbital physics taken into account.

\section*{Acknowledgements}
We would like to thank Philip W.~Phillips, Laura H.~Greene, and Wan Kyu Park, for helpful
discussions. This work is supported by the Center for Emergent
Superconductivity, a DOE Energy Frontier Research Center, Grant
No.~DE-AC0298CH1088, and W.L. is supported in part by the National Science Foundation Grant No.
DMR-1104386.

\end{document}